\documentclass[aps,prd,twocolumn,superscriptaddress,
floatfix,nofootinbib,showpacs]{revtex4}
\usepackage[dvips]{graphicx}
\usepackage{xspace}
\usepackage{latexsym}
\usepackage{amsmath,amssymb}
\usepackage{times}
\usepackage{mathptm}

\begin{document}

\title{Search for Proton Decay via $p \rightarrow \nu K^{+}$ using 260 kiloton$\cdot$year data of Super-Kamiokande}

\newcommand{\AFFicrr}{\affiliation{Kamioka Observatory, Institute for Cosmic Ray Research, University of Tokyo, Kamioka, Gifu 506-1205, Japan}}
\newcommand{\AFFkashiwa}{\affiliation{Research Center for Cosmic Neutrinos, Institute for Cosmic Ray Research, University of Tokyo, Kashiwa, Chiba 277-8582, Japan}}
\newcommand{\AFFipmu}{\affiliation{Kavli Institute for the Physics and
Mathematics of the Universe (WPI), Todai Institutes for Advanced Study,
University of Tokyo, Kashiwa, Chiba 277-8582, Japan }}
\newcommand{\AFFmad}{\affiliation{Department of Theoretical Physics, University Autonoma Madrid, 28049 Madrid, Spain}}
\newcommand{\AFFbu}{\affiliation{Department of Physics, Boston University, Boston, MA 02215, USA}}
\newcommand{\AFFbnl}{\affiliation{Physics Department, Brookhaven National Laboratory, Upton, NY 11973, USA}}
\newcommand{\AFFuci}{\affiliation{Department of Physics and Astronomy, University of California, Irvine, Irvine, CA 92697-4575, USA }}
\newcommand{\AFFcsu}{\affiliation{Department of Physics, California State University, Dominguez Hills, Carson, CA 90747, USA}}
\newcommand{\AFFcnm}{\affiliation{Department of Physics, Chonnam National University, Kwangju 500-757, Korea}}
\newcommand{\AFFduke}{\affiliation{Department of Physics, Duke University, Durham NC 27708, USA}}
\newcommand{\AFFfukuoka}{\affiliation{Junior College, Fukuoka Institute of Technology, Fukuoka, Fukuoka 811-0295, Japan}}
\newcommand{\AFFgmu}{\affiliation{Department of Physics, George Mason University, Fairfax, VA 22030, USA }}
\newcommand{\AFFgifu}{\affiliation{Department of Physics, Gifu University, Gifu, Gifu 501-1193, Japan}}
\newcommand{\AFFuh}{\affiliation{Department of Physics and Astronomy, University of Hawaii, Honolulu, HI 96822, USA}}
\newcommand{\AFFkanagawa}{\affiliation{Physics Division, Department of Engineering, Kanagawa University, Kanagawa, Yokohama 221-8686, Japan}}
\newcommand{\AFFkek}{\affiliation{High Energy Accelerator Research Organization (KEK), Tsukuba, Ibaraki 305-0801, Japan }}
\newcommand{\AFFkobe}{\affiliation{Department of Physics, Kobe University, Kobe, Hyogo 657-8501, Japan}}
\newcommand{\AFFkyoto}{\affiliation{Department of Physics, Kyoto University, Kyoto, Kyoto 606-8502, Japan}}
\newcommand{\AFFumd}{\affiliation{Department of Physics, University of Maryland, College Park, MD 20742, USA }}
\newcommand{\AFFmit}{\affiliation{Department of Physics, Massachusetts Institute of Technology, Cambridge, MA 02139, USA}}
\newcommand{\AFFmiyagi}{\affiliation{Department of Physics, Miyagi University of Education, Sendai, Miyagi 980-0845, Japan}}
\newcommand{\AFFnagoya}{\affiliation{Solar Terrestrial Environment Laboratory, Nagoya University, Nagoya, Aichi 464-8602, Japan}}
\newcommand{\AFFpol}{\affiliation{National Centre For Nuclear Research, 00-681 Warsaw, Poland}}
\newcommand{\AFFsuny}{\affiliation{Department of Physics and Astronomy, State University of New York at Stony Brook, NY 11794-3800, USA}}
\newcommand{\AFFniigata}{\affiliation{Department of Physics, Niigata University, Niigata, Niigata 950-2181, Japan }}
\newcommand{\AFFokayama}{\affiliation{Department of Physics, Okayama University, Okayama, Okayama 700-8530, Japan }}
\newcommand{\AFFosaka}{\affiliation{Department of Physics, Osaka University, Toyonaka, Osaka 560-0043, Japan}}
\newcommand{\AFFseoul}{\affiliation{Department of Physics, Seoul National University, Seoul 151-742, Korea}}
\newcommand{\AFFshizuokasc}{\affiliation{Department of Informatics in
Social Welfare, Shizuoka University of Welfare, Yaizu, Shizuoka, 425-8611, Japan}}
\newcommand{\AFFskk}{\affiliation{Department of Physics, Sungkyunkwan University, Suwon 440-746, Korea}}
\newcommand{\AFFtohoku}{\affiliation{Research Center for Neutrino Science, Tohoku University, Sendai, Miyagi 980-8578, Japan}}
\newcommand{\AFFtokyo}{\affiliation{The University of Tokyo, Bunkyo, Tokyo 113-0033, Japan }}
\newcommand{\AFFtokai}{\affiliation{Department of Physics, Tokai University, Hiratsuka, Kanagawa 259-1292, Japan}}
\newcommand{\AFFtit}{\affiliation{Department of Physics, Tokyo Institute
for Technology, Meguro, Tokyo 152-8551, Japan }}
\newcommand{\AFFtsinghua}{\affiliation{Department of Engineering Physics, Tsinghua University, Beijing, 100084, China}}
\newcommand{\AFFwarsaw}{\affiliation{Institute of Experimental Physics, Warsaw University, 00-681 Warsaw, Poland }}
\newcommand{\AFFuw}{\affiliation{Department of Physics, University of Washington, Seattle, WA 98195-1560, USA}}

\AFFicrr
\AFFkashiwa
\AFFmad
\AFFbu
\AFFbnl
\AFFuci
\AFFcsu
\AFFcnm
\AFFduke
\AFFfukuoka
\AFFgifu
\AFFuh
\AFFkek
\AFFkobe
\AFFkyoto
\AFFmiyagi
\AFFnagoya
\AFFsuny
\AFFokayama
\AFFosaka
\AFFseoul
\AFFshizuokasc
\AFFskk
\AFFtokai
\AFFtokyo
\AFFipmu
\AFFtsinghua
\AFFuw

\author{K.~Abe}
\author{Y.~Hayato}
\AFFicrr
\AFFipmu
\author{K.~Iyogi}
\AFFicrr 
\author{J.~Kameda}
\author{M.~Miura} 
\author{S.~Moriyama} 
\author{M.~Nakahata} 
\author{S.~Nakayama}
\author{R.~A.~Wendell} 
\author{H.~Sekiya} 
\author{M.~Shiozawa} 
\author{Y.~Suzuki} 
\author{A.~Takeda} 
\AFFicrr
\AFFipmu
% ...

\collaboration{The Super-Kamiokande Collaboration}
\noaffiliation

\date{\today}

\begin{abstract}

We have searched for proton decay via $p \rightarrow \nu K^{+}$ using
Super-Kamiokande data from April 1996 to February 2013, 260
kiloton$\cdot$year exposure in total.  No evidence for this proton
decay mode is found. A lower limit of the proton lifetime is set to
$\tau/B(p \rightarrow \nu K^{+}) > 5.9 \times 10^{33}$ years at 90\%
confidence level.

\end{abstract}

\maketitle
\newpage

\section{INTRODUCTION}

The standard model of particle physics, based on $SU(3)$ for the
strong interaction and the unification of $SU(2)\times U(1)$ for the
electroweak interaction, has been successful in accounting for many
experimental results. However, the standard model offers no guidance
on the unification of the strong and electroweak forces, and has many
other open questions. Various attempts have been made to resolve the
shortcomings by unifying the strong and electroweak interactions in a
single larger gauge group, i.e. a Grand Unified Theory
(GUT)~\cite{guts}. GUTs are motivated by the apparent convergence of
the running couplings of the strong, weak, and electromagnetic forces
at a high energy scale ($10^{15}-10^{16}$\,GeV). Energy scales this
large are out of the reach of accelerators but may be probed by
virtual processes such as those that govern particle decay. A general
feature of GUTs is the instability of nucleons by baryon number
violating decay. Nucleon decay experiments are direct experimental
tests of the general idea of grand unification.

In GUTs, nucleon decay can proceed via exchange of a massive gauge
boson between two quarks. The favored gauge-mediated decay mode in
many GUTs is $p \rightarrow e^{+} \pi^{0}$. In the minimal SU(5) GUT,
the predicted proton lifetime to $e^+\pi^0$ is $10^{31\pm 1}$ years,
which has been ruled out by experimental results from IMB~\cite{imb},
Kamiokande~\cite{kam}, and Super-Kamiokande~\cite{skepi0}. GUT models
incorporating supersymmetry~\cite{susy} (SUSY-GUTs) raise the GUT
scale~\cite{Marciano:1981un}, suppressing the decay rate of $p
\rightarrow e^{+} \pi^{0}$, thereby allowing compatibility with the
experimental limit. However, SUSY-GUTs introduce dimension five
operators that enable the mode $p \rightarrow \overline{\nu} K^{+}$ to
have a high branching fraction and short partial lifetime~\cite{d5}. In
the SUSY SU(5) GUT with minimal assumptions and TeV scale SUSY
particles, the partial proton lifetime to $\overline{\nu} K^+$ is less
than $10^{31}$ years~\cite{Murayama:2001ur}, which has also been
excluded by previously published experimental
constraints~\cite{imb,kam,Hayato:1999az}. Non-minimal SUSY SU(5)
GUTs~\cite{nonminimalsusysu5} or SUSY GUTs based on
SO(10)~\cite{susyso10} have been constructed that evade this limit, yet
still predict partial lifetimes in the range $10^{32}$ to $10^{35}$
years, with some particular models that require the lifetime be less
than a few times $10^{34}$ years. The low ends of the lifetime
predictions by these models are probed by this experimental search.

In this paper, our search is for the two-body decay of proton to a
$K^+$ and a neutrino. In most models, ($B-L$) is conserved and the
final state contains an anti-neutrino; however we do not detect the
neutrino and cannot distinguish $p \rightarrow \nu K^+$ from $p
\rightarrow \overline{\nu} K^+$, nor can we determine the flavor ($e$,
$\mu$, or $\tau$) of the neutrino. In fact, our search can be applied
to any nearly massless neutral final-state particle such as a gravitino or axino.

The Super-Kamiokande collaboration published a search for $p
\rightarrow \nu K^{+}$ with 91.7 kton$\cdot$years exposure of the
first phase of the experiment, and set a partial lifetime limit
$\tau(p \rightarrow \nu K^+) > 2.3 \times 10^{33}$ years~\cite{kk}. In
this paper, we refine the analysis and update the search with 2.8
times greater detector exposure including later phases of the
experiment.

\section{SUPER-KAMIOKANDE DETECTOR}

Super-Kamiokande~\cite{skdet} is a large water Cherenkov detector. It
is an upright cylinder in shape, 39\,m in diameter and 40\,m in
height, and it contains 50\,kton of pure water. It lies about 1,000\,m
underneath the top of Mt.~Ikenoyama (2,700\,m water equivalent
underground) to reduce cosmic ray background.  The detector is
optically separated into two regions: inner detector (ID) and outer
detector (OD). Cherenkov light in the ID is detected by 20-inch
PMTs~\cite{pmt} facing inward, evenly covering the cylindrical inner
surface. Cherenkov light from penetrating particles, usually cosmic
ray muons or exiting muons, is detected by 8-inch PMTs facing outward.
The fiducial volume is defined as a cylindrical volume with surfaces 2
meters inwards from the ID PMT plane. The fiducial mass is
22.5\,ktons, corresponding to $7.5\times 10^{33}$ protons.

Super-Kamiokande started observation in April 1996 with 11,146 PMTs
which covered 40\% of the ID surface. The observation was continued
until July 2001, with 1489.2 live days, or equivalently,
91.7\,kton$\cdot$years. This period is called Super-Kamiokande-I
(SK-I). After an accident in 2001, about half of the ID PMTs were lost
and the detector was reconstructed with 5,182 ID PMTs uniformly
distributed over the cylindrical surface decreasing photo coverage to
19\%. The PMTs were thereafter enclosed in acrylic and FRP cases. The
period from December 2002 until October 2005, corresponding to 798.6
live days (49.2\,kton$\cdot$years), is called SK-II. After production
and installation of replacement 20-inch PMTs, the photo coverage was
recovered to 40\% in 2006. The period between July 2006 and September
2008, corresponding to 518.1 live days (31.9\,kton$\cdot$years), is
defined as SK-III. In the summer of 2008, we upgraded our electronics
with improved performance including a data acquisition that records
all PMT hit information without dead time~\cite{elec}. This has been the
configuration of the detector since September 2008; it is called
SK-IV. In this paper, we use data until February 2013, corresponding
to 1417.4 live days (87.3\,kton$\cdot$years). Table~\ref{data}
summarizes the data sets used for the proton decay search in this
paper.

\begin{table}
  \begin{center}
    \begin{tabular}{lrrcl}
      \hline \hline
                    & Live days &  kton$\cdot$yr  &  Coverage & Note         \\
      \hline 
            SK-I    & 1489.2    &  91.7           &  40\%     &               \\
            SK-II   &  798.6    &  49.2           &  19\%     &  Half PMT density \\
            SK-III  &  518.1    &  31.9           &  40\%     &                \\
            SK-IV   & 1417.4    &  87.3           &  40\%     & New readout electronics\\
      \hline \hline
    \end{tabular}
  \caption{\protect \small 
Summary of data sets that are used in this paper.
}
  \label{data}
  \end{center}
\end{table}

The trigger to record an event is based on the coincidence of the
number of hit PMTs exceeding a threshold. For SK I-III, the trigger
was implemented in hardware using a signal proportional to the number
of hit PMTs produced by each front-end electronics module. For SK-IV,
the trigger is implemented in software.  The trigger threshold is less
than 10\,MeV for all SK periods, and the trigger efficiency for this
proton decay mode is 100\%\footnote{For the case of prompt gamma tag
  (Method 1), with gamma energy typically 6 MeV, the muon from $K^+$
  decay can provide the event trigger.}.

The charge and timing of the PMTs are calibrated using various
calibration sources~\cite{calib}. The timing resolution of the 20-inch
PMT is about 2.1\,nsec at the single photo-electron level. The PMT
response, water quality, and reflections from the detector wall are
tuned in the SK detector simulation program using injected light as
well as various control data samples such as cosmic ray muons.

\section{SIMULATION}

To determine selection criteria for the proton decay search, and
to estimate efficiencies and background rates, we use proton decay and
atmospheric neutrino Monte Carlo (MC) simulations. Because the
configuration of the detector is different in SK-I through IV, we
generated MC samples for each period. Proton decay MC samples with
50,000 events are generated in an oversized volume which is 1 meter
outside the fiducial volume boundary, and therefore 1 meter from the
detector wall. This allows us to include event migration near the
fiducial boundary in our estimates. The selection efficiency is
defined as the number of events fulfilling all requirements divided by
number of generated events in the fiducial volume. The MC equivalent
of 500 years of atmospheric neutrino exposure are generated for each
period. These atmospheric neutrino MC samples are used for our studies 
of neutrino oscillations~\cite{nuosc}. Because the background rates for
the proton decay studied in this paper are small (less than one event
for the entire exposure for two of the analysis techniques), these
large MC background samples provide fewer than 40 atmospheric
neutrino events that survive the proton decay selection criteria.

\subsection{Proton Decay}

A water molecule contains two free protons and eight protons bound
in the oxygen nucleus.  In the decay of a free proton, the $\nu$ and
the $K^+$ are emitted opposite each other with momenta of
339\,MeV/$c$.  In the case of proton decay in oxygen, Fermi momentum,
correlation with other nucleons, nuclear binding energy, and
kaon-nucleon interactions are taken into account as described below.

\begin{figure}[htbp]
\begin{center}
  \includegraphics[width=8cm,clip]{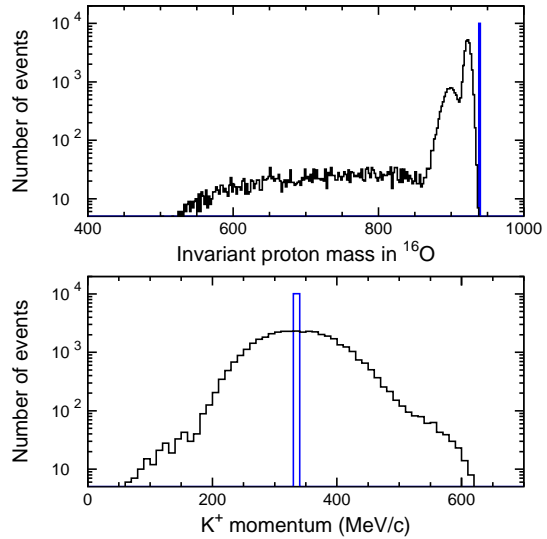}
  \caption
  {\protect \small (color online) The upper figure shows the decaying
    proton mass distribution in $^{16}$O and the lower figure shows
    the $K^+$ momentum distribution from the simulation of $p
    \rightarrow \nu K^+$. In the upper figure, the single-bin
    histogram shows the free proton case and the broad histogram shows
    the bound proton case.  The rightmost peak in the bound proton
    case corresponds to the $p$-state, located slightly lower than the
    proton mass by 15.5\,MeV of binding energy; the second rightmost
    peak is the $s$-state (39\,MeV in binding energy). The correlated
    nucleon decay makes the longer tail in the lower mass region. In
    the lower figure, the single-bin histogram shows the free proton
    case (339\,MeV/c) and the broad histogram shows the bound proton case
    which is smeared by Fermi motion.}
  \label{kpmom}
\end{center}
\end{figure}

We use the Fermi momentum and nuclear binding energy measured by
electron-$^{12}$C scattering~\cite{fermi}.  Nuclear binding energy is
taken into account by modifying the proton mass. Ten percent of
decaying protons are estimated to have wave functions which are 
correlated with other nucleons within the
nucleus~\cite{corrdcy}. These correlated decays cause the total
invariant mass of the decay products to be smaller than the proton
mass because of the momentum carried by the correlated nucleons.
Figure~\ref{kpmom} shows the invariant mass of the products of the
decaying proton, $K^+$ and $\nu$ and the resulting kaon momenta after
the simulation of the proton decay for both bound and free protons.
Correlated decays produce the broad spectrum below about
850\,MeV/$c^2$. In our experiment, the kaon momentum is unobserved
because the kaon is always produced below the Cherenkov threshold of
749\,MeV/$c$ in water. The majority of $K^+$ (89\%) are stopped in
water and decay at rest. We search for $K^+$ decay at rest into $\mu^+
\nu_\mu$ (64\% branching fraction) and $\pi^+ \pi^0$ (21\% branching
fraction).

The position of the decaying proton in $^{16}O$ is calculated
according to the Woods-Saxon nuclear density model~\cite{woodsax}.
The kaon nucleon interactions which are considered include elastic
scattering and inelastic scattering via charge exchange. The type of
interaction is determined using the calculated mean free path
~\cite{neut}. For kaons, whose momenta are described by
Fig.~\ref{kpmom}, the probability of charge exchange for $K^{+}$ in $p
\rightarrow \nu K^{+}$ is 0.14\%.

If a nucleon decays in the oxygen nucleus, the remaining nucleus can
be left in an excited state from which it promptly de-excites by the
emission of gamma rays. The prompt gamma emission processes are
simulated based on reference~\cite{ejiri}. The dominant gamma ray is
6.3\,MeV from the $p_{3/2}$ state with 41\% branching fraction. The
probabilities of $\gamma$ emission in this simulation are summarized
in Table~\ref{emission}. Other states emitting low energy gamma rays
are averaged and assigned 3.5\,MeV $\gamma$ emission. Nuclear decay
into states that emit protons or neutrons and nuclear decay into the
ground state are taken to have no $\gamma$ ray emission.

\begin{table}
  \setlength{\tabcolsep}{4pt} 
  \begin{center}
    \begin{tabular}{lrr}
      \hline \hline
              State &  Energy of $\gamma$    &  Probability  \\
      \hline 
          $p_{3/2}$ & 6.3\,MeV     &  41\%          \\
          $p_{3/2}$ & 9.9\,MeV     &   3\%          \\
          $s_{1/2}$ & 7.03\,MeV    &   2\%          \\
          $s_{1/2}$ & 7.01\,MeV    &   2\%          \\
          others    & 3.5\,MeV     &  16\%          \\
      \hline
          Other than $\gamma$ emission & &       \\
      \hline
          $p$/$n$ emission &  - &  11\%          \\
          ground state &  -     & 25\%          \\
      \hline \hline
    \end{tabular}
  \caption{\protect \small 
Summary of probabilities of nuclear $\gamma$ ray emissions at the de-excitation 
of the remaining nucleus.
}
  \label{emission}
  \end{center}
\end{table}

\subsection{Atmospheric Neutrinos}

\begin{figure}[htbp]
\begin{center}
  \hspace*{-0.5cm}\includegraphics[width=8cm]{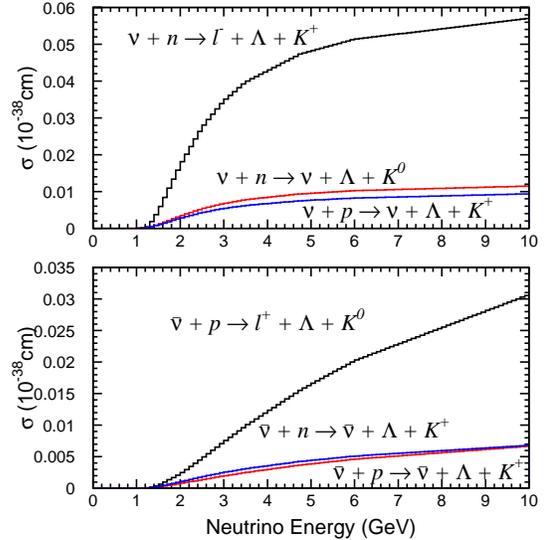}
  \caption
  {\protect \small (color online) Cross sections of the single K-meson
     productions via resonances calculated by NEUT. Upper plots show
     neutrino interactions and lower show anti-neutrino interactions.}
  \label{kcrs}
\end{center}
\end{figure}

The SK standard atmospheric neutrino MC used in the previous neutrino
oscillation analyses~\cite{nuosc} and proton decay
searches~\cite{skepi0,kk,numeson, muk0} is used in this analysis. It
is based on the Honda atmospheric neutrino flux~\cite{honda} and
NEUT~\cite{neut} neutrino-nucleus interaction model.  Some neutrino
interactions which produce K mesons via resonances could be potential
background sources for $p \rightarrow \nu K^{+}$ search.  Cross
sections of the single meson production via resonances are calculated
based on Rein and Sehgal's theory \cite{rein-sehgal}. In NEUT, the
neutrino reactions:

\begin{align}
\nu~n  &\rightarrow l^{-}~\Lambda~K^{+} \notag \\
\nu~n &\rightarrow \nu~\Lambda~K^{0} \notag \\
\nu~p &\rightarrow \nu~\Lambda~K^{+} \notag \\
\overline{\nu}~p &\rightarrow l^{+}~\Lambda~K^{0} \notag \\
\overline{\nu}~n &\rightarrow \overline{\nu}~\Lambda~K^{0} \notag \\
\overline{\nu}~p &\rightarrow \overline{\nu}~\Lambda~K^{+} \notag
\end{align} 

\noindent are taken into account assuming the same cross section both
for $\nu_{e}$ and $\nu_{\mu}$. The differential cross sections are
shown in Fig.~\ref{kcrs}.

We simulate
propagation of the produced particles and Cherenkov light in water by
custom code based on GEANT3~\cite{geant}. The propagation of charged
pions in water is simulated by custom code based on Ref.~\cite{skpimc}
for less than 500\,MeV/$c$ and by GCALOR~\cite{calor} for more than
500\,MeV/$c$.

The equivalent of 500 years of SK atmospheric neutrino data is
simulated for each SK run period. The generated atmospheric neutrino
samples are weighted to include the effect of $\nu_\mu$ disappearance
due to $\nu_\mu$-$\nu_\tau$ oscillation assuming $\Delta m^{2} = 2.5
\times 10^{-3}$ eV$^2$ and sin$^{2}2\theta=1.0$, ignoring the
appearance of $\nu_e$ or $\nu_\tau$ as a possible background. The
final background event rates for each period are normalized by the
observed total sub-GeV event rate.

\section{DATA SET, REDUCTION AND RECONSTRUCTION}

The vast majority of the triggered events are cosmic ray muons and low
energy backgrounds from the radioactivity of materials around the
detector wall. Several stages of data reduction were applied to the
events before proceeding to further detailed event reconstruction
processes. Details of the data reduction and reconstruction can be
found in \cite{nuosc}.

The fully contained (FC) data sample in the fiducial volume (FV) is
defined by the following cuts:
\begin{itemize}
\item number of hit PMTs in the largest OD hit cluster is less than 10 for SK-I and 16 for other period.
\item total visible energy is greater than 30\,MeV in ID
\item distance of the reconstructed vertex from the ID PMT surface is
  greater than 2\,meters (corresponding to 22.5\,kton of water volume)
\end{itemize}

\noindent The rate of FCFV events is about 8 events per day. The
contamination of events other than atmospheric neutrinos is estimated
to be less than 1\% and composed of cosmic rays that evaded the OD
veto and events caused by flashing PMTs.

Reconstruction algorithms are applied to the events remaining after
the reduction process to determine the event vertex, the number of
Cherenkov rings, the particle type of each ring, the momentum assigned
to each ring, and number of Michel electrons. As a first step, the
event vertex is defined as the point at which the timing distribution,
after subtraction of the calculated time of flight of the photon from
the vertex (TOF subtraction), has the sharpest peak.  The dominant
ring direction is determined from the charge distribution as a
function of angle. Then other ring candidates are searched for using
the Hough transform method~\cite{hogh}, a technique for extracting a
particular shape from an image, assuming all particles are generated
in one vertex. Each ring candidate is tested against a likelihood function 
to remove fake rings before determining the final number of rings. Each ring
is classified as $e$-like (showering type as from $e^{\pm,}, \gamma$)
or $\mu$-like (non-showering type) based on a likelihood using the
ring pattern and Cherenkov opening angle for single ring case, and only
ring pattern for multi-ring case.

Michel electrons from the decay of the $\mu^{\pm}$ are tagged by
searching for clusters of in-time hits after the primary event. During
the SK-I to SK-III periods, there was an impedance mismatch between
cables and electronics which caused a signal reflection at 1000~ns
after the main event. The time period between 800\,ns and 1200\,ns
from the primary events was excluded for the decay electron search due
to this signal reflection.  For SK-IV, the new electronics have better
impedance matching to avoid signal reflection, and no such exclusion
is required. As a result, the tagging efficiency of decay electrons
from $\mu^{+}$ with momentum of 236\,MeV/c has been improved from 85\% 
(SK-I, II, and III) to 99\% (SK-IV), which improves the selection 
efficiency for $p \rightarrow \nu K^{+}$.

The momentum for each ring is decided from the charge spatially inside
of 70$^\circ$ from the ring direction and temporally from $-$50\,nsec
to +250\,nsec around the TOF-subtracted main event peak.  The number of
photoelectrons from each PMT are corrected by light attenuation in
water, PMT acceptance, dark noise, time variation of gain, and track
length in the case of $\mu$-like rings.  If a Michel electron is
within 250\,nsec of the parent particle, the time window for momentum
determination is shortened and the charge sum is corrected from the
nominal +250\,nsec case. The momentum scale is checked by cosmic ray
muons, Michel electrons from the cosmic ray muons which stop in the
inner detector, and also the invariant mass distributions of $\pi^0$
events produced in atmospheric neutrino interactions. The uncertainty
on the momentum scale is less than 3\% for the entire period.

An additional precise vertex fitter is applied only for single-ring
events. The expected charge for each PMT is calculated using the
result of particle identification ($e$-like or $\mu$-like) and using
the momentum estimate. The expected charge is compared with the
observed charge by varying the vertex along the particle
direction. The estimated vertex resolution for FC single-ring sub-GeV
events is about 30\,cm.

\section{ANALYSIS}

If a proton decays by $p \rightarrow \nu K^{+}$, the $K^{+}$ itself is
not visible in a water Cherenkov detector since its velocity is below
Cherenkov threshold. However, the $K^{+}$ can be identified by its
decay products in the decay modes $K^+ \rightarrow \mu^{+} \nu$ and
$K^+ \rightarrow \pi^{+} \pi^{0}$. Being two-body decay processes, the
daughter particles have monochromatic momentum in the $K^+$ rest
frame: 236\,MeV/$c$ for $\mu^+ \nu_\mu$ and 205\,MeV/$c$ for $\pi^+
\pi^0$. There are three established methods for the $p \rightarrow \nu
K^{+}$ mode search~\cite{Hayato:1999az}: (Method 1) since the $\gamma$
ray is promptly emitted at the time of $K^+$ production, look for
single muon events with a de-excitation $\gamma$ ray just before the
time of the muon; (Method 2) search for an excess of muon events with
a momentum of 236\,MeV/$c$ in the momentum distribution; and (Method
3) search for $\pi^{+}\pi^{0}$ events with a momentum of 205\,MeV/$c$.

\subsection{Method 1: $K^+ \rightarrow \mu^{+} \nu$, tag by prompt gamma ray}

\begin{figure}[htbp]
\begin{center}
  \hspace*{-0.5cm}\includegraphics[width=8cm,clip]{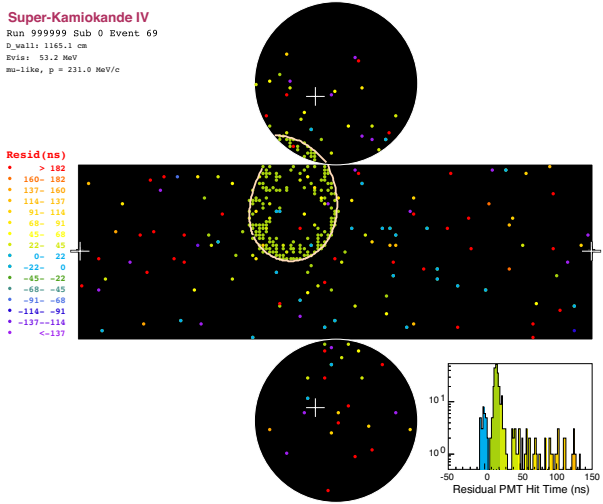}
  \caption
  {\protect \small (color online) An example graphical event display
    of a simulated proton decay passing all of the criteria for Method
    1. The single Cherenkov ring was produced by the muon from kaon
    decay and fit with momentum 231 MeV/$c$. The color of the hit PMTs
    represents the residual hit time after subtracting the
    time-of-flight of Cherenkov light in water from the vertex to the
    PMT. The hit PMTs associated with a 6 MeV prompt gamma are colored
    cyan. The decay electron was detected after the displayed event.}
  \label{evd-gammatag}
\end{center}
\end{figure}

\begin{figure}[htbp]
\begin{center}
  \includegraphics[width=8cm,clip]{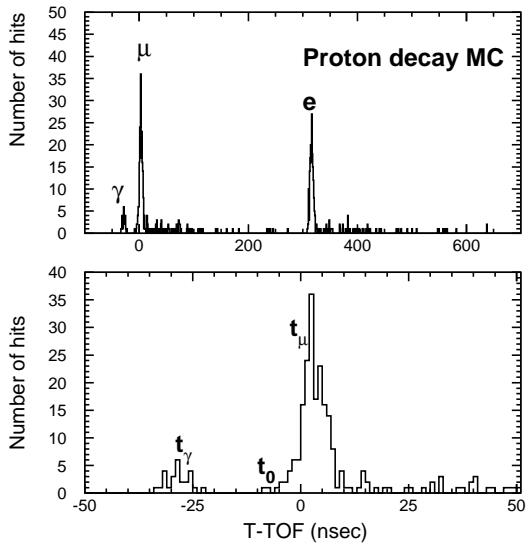}
  \caption
  {\protect \small Number of hits versus hit time for a typical proton
    decay MC event selected by Method 1. In the upper figure, three
    hit clusters due to prompt $\gamma$, $\mu$, and the Michel
    electron can be seen in order. The lower figure is expanded around
    the $\mu$-cluster and $t_{\gamma}$, $t_{0}$, and $t_{\mu}$ are
    shown as a demonstration.  }
  \label{t-sample}
\end{center}
\end{figure}

If this proton decay happens, the Super-K detector should observe a
single $\mu$-like ring preceded by PMT hits due to a nuclear
deexcitation gamma ray. Figure~\ref{evd-gammatag} shows a graphical
event display of the PMT hit pattern for an example event, as
a Monte Carlo simulation of the proton decay. A prompt gamma ray, a muon, and a Michel
electron peak should be observed in order, as seen in another example
in Fig.~\ref{t-sample}, where the time of each particle is labeled as
used in the analysis. To search for events tagged by the prompt gamma
ray, single-ring $\mu$-like events are selected by requiring the
following criteria:

\begin{itemize}
\item[(A-1)] a fully contained event with one non-showering
  ($\mu$-like) ring,
\item[(A-2)] there is one Michel decay electron, 
\item[(A-3)] the reconstructed muon momentum is between 215 and 260\,MeV/$c$, 
\item[(A-4)] the distance between the vertices of the muon and the
  Michel electron is less than 200\,cm,
\item[(A-5)] the TOF-subtracted timing distribution for the muon
  vertex is required to have a minimum goodness-of-fit ($>$ 0.6),
\item[(A-6)] the pattern of the single non-showering ring is more
  likely to be a muon than a proton: $L_{pr}-L_{\mu} < $ 0, $L_{pr}$,
  $L_{\mu}$ are likelihood functions assuming a proton and a muon,
  which are described later,
\item[(A-7)] gamma hits are found: $8 < N_{\gamma} < 60$ for SK-I, III, and VI,
            $4 < N_{\gamma} < 30$ for SK-II
\item[(A-8)] the time difference from the gamma tag to the kaon decay
  is consistent with the kaon lifetime: $t_{\mu}-t_{\gamma} <
  75$\,nsec,
\end{itemize}

The cut criteria (A-4) and (A-5) are applied to reject events with a
high momentum recoil proton (above Cherenkov threshold) accompanied by
an invisible muon or charged pion (below the Cherenkov threshold)
which produces a Michel electron in its decay chain. Since the particle
type of the single non-showering Cherenkov ring is assumed to be that
of a muon, the vertex accuracy is worse when the Cherenkov ring is
from a recoil proton. The inaccurate vertex determination causes
incorrect TOF subtraction of the Cherenkov light. As a result, recoil
protons may create a false peak in the time distribution of hit PMTs,
which can fake a prompt gamma ray. The event may include Michel
electrons from the decay of the invisible muon, but the distance
between the misreconstructed vertex and the Michel electron is typically large.

\begin{figure}[htbp]
\begin{center}
  \includegraphics[width=8cm,clip]{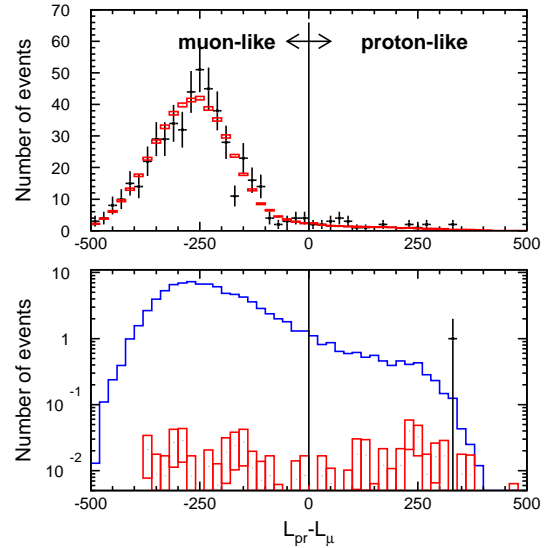}
  \caption
  {\protect \small (color online) The likelihood distribution to
    separate $\mu$ and proton.  The negative region is $\mu$-like
    and positive is proton-like.  In the upper figure, the atmospheric
    $\nu$ MC (box) is compared with data merging SK-I/III/IV (dot)
    requiring cuts (A-1) through (A-5), and they are in good
    agreement. The lower figure shows the same distributions after
    $\gamma$-tagging cuts; the remaining background is reduced by the
    likelihood cut.}
  \label{p-pid}
\end{center}
\end{figure}

The proton identification criterion (A-6) is a refinement to the
methods in our previous paper for rejecting single proton ring events. 
It is used for reduction of single
proton ring events. The algorithm~\cite{protonid} makes a likelihood
function assuming a muon ($L_{\mu}$) and a proton ($L_{pr}$) by
using the Cherenkov angle and the width of the Cherenkov
ring. Figure~\ref{p-pid} shows the likelihood function. The upper
figure is for the sample requiring cuts (A-1) through (A-5).
Data and MC agree well. The lower
figure is the same distribution after applying cuts (A-7) and (A-8) additionally.
Background events are efficiently reduced by (A-6).

After cuts (A-1) through (A-6), a distribution of hits ($N$) vs. time 
after TOF subtraction ($t$) is made. 
To search for
the prompt gamma ray, three quantities of time must be defined. The
first is $t_{\mu}$, which represents the time associated with the
detection of the muon, or equivalently, the decay of the kaon. The second
is $t_{0}$, which is the start time to search in the past time
distribution of hits to find the prompt gamma ray hits. The third is
$t_{\gamma}$, which is the associated time of the gamma ray
detection. PMTs outside of a 50$^\circ$ cone with respect to the muon
direction are masked and $t_{\mu}$ is defined as the time where
$dN/dt$ is maximum. The signal of the gamma ray is so tiny, compared to
the muon, that it can easily be hidden by muon hits. To avoid this, the
gamma finding is started earlier than the muon hits. To determine
$t_{0}$, $dN/dt$ is calculated from the muon peak time into the
past. Muon hits are dominant while $dN/dt\,>\,0$, and $t_{0}$ is
defined at the point where $dN/dt$ changes to less than or equal to
0. Then, in the $N$-versus-$t$ distribution, a time window with
12\,nsec width is slid backward from $t_{0}$. The associated time of
the gamma ray candidate, $t_{\gamma}$, is defined as the middle of the
time window where the number of hits in the window is maximum,
$N_{\gamma}$.

\begin{figure}[htbp]
\begin{center}
  \includegraphics[width=8cm,clip]{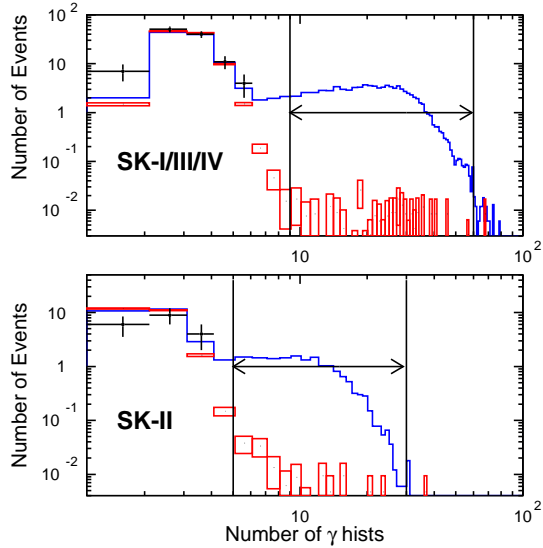}
  \caption
  {\protect \small (color online) Number of $\gamma$ ray hit
    distributions. The upper figure shows sum of SK-I, III, and IV
    which have 40\% photo coverage, the lower figure corresponds to
    SK-II with 19\% photo coverage. Dots, squares, and histogram
    correspond to data, atmospheric $\nu$ MC normalized to livetime of data,
     and proton decay MC with
    arbitrary normalization, respectively. The signal regions are
    indicated by arrows.  The peaks at small numbers of hits are due to dark
    hits of the PMTs. }
  \label{ngam}
\end{center}
\end{figure}

Figure~\ref{ngam} shows the $N_{\gamma}$ distribution for SK-I, III,
and IV in the upper figure and SK-II in the lower figure after all
cuts except (A-7).  An arrow in the figure shows the signal region,
i.e. with cut (A-7) applied; there are no data in this region.

Table~\ref{reduction-gam} show background rate per Megaton$\cdot$year and
expected events estimated by atmospheric $\nu$ MC, observed events in data,
and efficiencies evaluated by the proton decay MC for each reduction step. 
SK-I, III, and IV which have 40\% photo-coverage are merged and results of SK-II
with 19\% photo-coverage are separately shown.

\begin{table*} 
\setlength{\tabcolsep}{7pt} 
\begin{center}
\begin{tabular}{r|rrrr|rrrr}
\hline \hline
           \multicolumn{1}{c}{} &
           \multicolumn{4}{c}{SK-I/III/IV} &
           \multicolumn{4}{c}{SK-II} \\           
Criterion & Bkg. Rate & Exp. Bkg. & Data & Signal Eff. &
             Bkg. Rate & Exp. Bkg & Data & Signal Eff. \\             
\hline
A-1 & 35240.8 & 7432.3 & 7497 & 0.575
    & 34910.6 & 1717.6 & 1712 & 0.566 \\     
A-2 & 24865.7 & 5244.2 & 5240 & 0.520
    & 22239.7 & 1094.2 & 1051 & 0.473 \\     
A-3 & 2496.6  & 526.5  & 531  & 0.494 
    & 2161.0  & 106.3  & 91   & 0.440 \\        
A-4 & 2443.7  & 515.4  & 520  & 0.485 
    & 2067.8  & 101.7  & 87   & 0.420 \\     
A-5 & 2400.3  & 506.2  & 514  & 0.479 
    & 2030.0  & 99.9   & 82   & 0.414 \\     
A-6 & 2302.7  & 485.6  & 488  & 0.436
    & 1931.5  & 95.0   & 78   & 0.368 \\ 
A-7 &  1.34   &  0.28  &  0   & 0.084
    &  5.84   &  0.29  &  0   & 0.063 \\     
A-8 &  1.11   &  0.24  &  0   & 0.084
    &  2.75   &  0.14  &  0   & 0.062 \\     
\hline \hline
\end{tabular}
\end{center}
\caption{\protect \small 
Event rates per Megaton$\cdot$year and expected numbers of event from
atmospheric $\nu$ MC, observed numbers of event in data, and signal
efficiencies estimated from proton decay MC, for each step. 
SK-I/III/IV with 40\% photo-coverage and SK-II with 19\% are shown separately.  
}
\label{reduction-gam}
\end{table*}   

The selection efficiencies, expected number of background, and
observed number of events are summarized in Table~\ref{result}. The
efficiency in SK-IV is higher than the other periods because the
tagging efficiency for Michel electrons has been improved thanks to
the new electronics described in the detector section. The total
expected background for 260~kton$\cdot$year exposure is 0.4 events,
and no events are observed. The dominant neutrino interaction in the
background expectation comes from $\Delta S =0$ kaon production
(48\%): $\nu p \rightarrow \nu K^{+} \Lambda$, where the $\Lambda$
decays to unobserved proton and $\pi^-$. If the neutrino interaction
is accompanied by de-excitation gamma rays, the event has the same
configuration as the proton decay signal. The second most prevalent
background is $\nu_\mu$ charged current quasi-elastic scattering
accompanied by de-excitation gamma rays (25\%).

The dominant systematic error for the selection of signal is
uncertainty in the de-excitation gamma ray emission
probabilities. They are estimated to be 15\% for the 6.3\,MeV gamma
ray and 30\% for the others~\cite{ejiri}, and they contribute 19\% to
the overall systematic uncertainty on the selection efficiency. The
other systematic uncertainties come from event reconstruction: energy
scale, particle ID, ring-counting, fiducial volume, water scattering
and attenuation parameters, and they range in size from 1\% to 3\%. In
total, 22\% is the systematic error of the selection. The uncertainty
in the background rate comes from atmospheric neutrino flux and the
cross section of neutrino interactions. The uncertainty in the
neutrino flux is conservatively estimated to be 20\%~\cite{honda}. By
changing the cross section of charged current quasi-elastic
scattering, neutral current elastic scattering, and single $\pi$
production interaction by $\pm$30\%, and the deep inelastic scattering
by $\pm$50\%, a 10\% uncertainty in background rate is found. The
total systematic error for the background in Method 1 is estimated to
be 25\%.

There were several improvements in our analysis since our last paper
about $p \rightarrow \nu K^{+}$ was published in 2005~\cite{kk}. As
described in the reconstruction section, the time window for hits used
to calculate momentum is changed if a Michel electron is closer than
250\,nsec from the parent particle.  This is a new algorithm which
prevents the overestimation of the momentum due to including PMT hits
from the Michel electron. Previously, the vertices of those events
tended to be misfit in the forward direction since the precise
fitter used the expected charge for each PMT based on an
overestimated momentum with larger Cherenkov angle. This resulted in
more TOF to be subtracted for hits backward of the particle direction
and, as a result, it sometimes made fake prompt $\gamma$ signals in
atmospheric $\nu$ interactions. The improvement of the momentum
calculation algorithm reduced the atmospheric $\nu$ background by a
factor of three and eliminated a candidate event in the SK-IV data
that would have survived based on the uncorrected algorithm. As a
result, the expected background for the gamma-tag method in SK-I is
reduced from the value in the previous paper 0.7 events, to 0.2 events
while maintaining signal efficiency. The new selection criterion (A-6)
further rejects 60\% of the atmospheric background (after all other
cuts) while losing only 8\% efficiency.  As a result, the expected
background for Method 1 is greatly reduced, finally to 0.08 events for
the SK-I period.

\subsection{Method 2: $K^+ \rightarrow \mu^{+} \nu_\mu$, mono-energetic muon search}

\begin{figure}[htbp]
\begin{center}
  \includegraphics[width=8cm,clip]{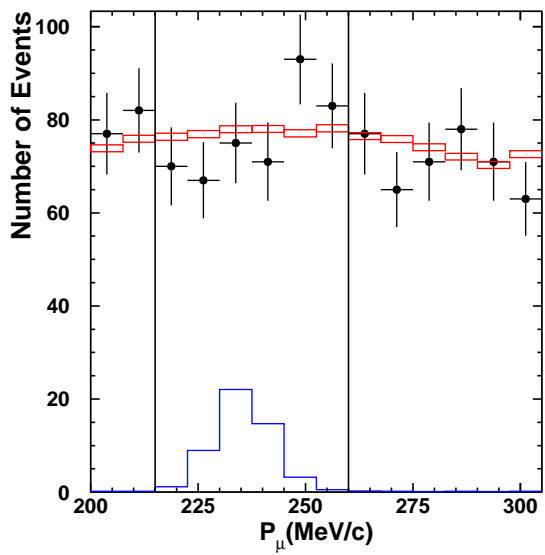}
  \caption
  {\protect \small (color online) Muon momentum distribution for
    260\,kton$\cdot$year. Dots, boxes, and histogram correspond to
    data, atmospheric $\nu$ MC, and proton decay MC, respectively. The
    data are fit by the background plus signal by free normalization.
    No excess above the expected background is observed. The
    normalization of the proton decay MC histogram shown is at the
    upper limit allowed by the fit.}
  \label{mumom}
\end{center}
\end{figure}

Since most of the $K^{+}$s stop in water, the monoenergetic $\mu^{+}$s
from kaon decays would lead to an excess peak in the muon momentum
distribution of atmospheric neutrino background. To avoid using the same
events as in Method 1, Method 2 requires all the criteria in Method 1
except: the requirements in momentum (A-3) are relaxed to allow a
spectrum fit, and the gamma hits (A-7) must not be present. We search
for an excess of muon events in the momentum distribution by fitting
the data with the proton decay signal expectation over the atmospheric
neutrino background events. The signal and background normalizations
are free parameters in the fit. Figure~\ref{mumom} shows the muon
momentum distribution of data from SK-I to SK-IV compared with MC. No
significant excess is observed in the signal region, defined by
vertical lines.

\subsection{Method 3: $K^+ \to \pi^{+}+\pi^{0}$ }

\begin{figure}[htbp]
\begin{center}
  \hspace*{-0.5cm}\includegraphics[width=8cm,clip]{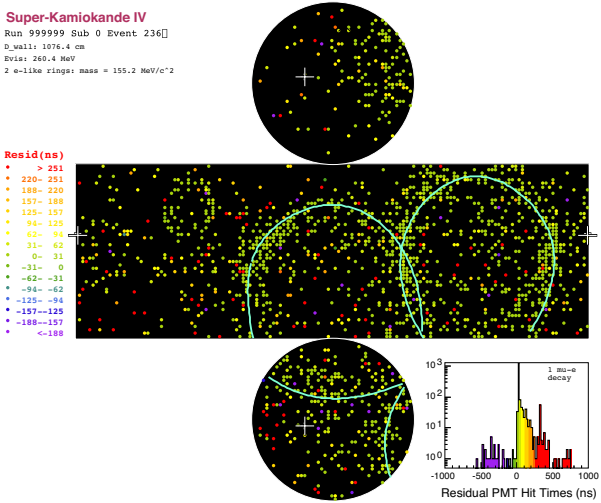}
  \caption
  {\protect \small (color online) An example graphical event display
    of a simulated proton decay passing all of the criteria for Method
    3. The two Cherenkov rings were produced by gamma rays from
    $\pi^0$ decay and reconstructed to an invariant mass of 155
    MeV/$c^2$ and momentum of 209 MeV/$c$. The color of the hit
    PMTs represents the residual hit time after subtracting the
    time-of-flight of Cherenkov light in water from the vertex to the
    PMT. The $\pi^+$ does not make a Cherenkov ring that is
    reconstructed, however PMT hits due to this particle are
    present opposite to the $\pi^0$ direction, visible in the upper left
    of the ID barrel region.}
  \label{evd-pipi0}
\end{center}
\end{figure}

In Method 3, $\pi^0$ events with a momentum of 205\,MeV/$c$ are
selected. An example event display is shown in Fig.~\ref{evd-pipi0}.
The $\pi^{0}$ decays into two photons; if the energy of one
photon is much smaller than the other, sometimes those events are
misidentified as a single-ring event. A special $\pi^0$ reconstruction
algorithm is used to search for proton decay candidates within the single-ring event sample. The
$\pi^+$ does not make a clear Cherenkov ring due to its low momentum.
However, hit activity in the opposite direction of the $\pi^0$, caused
by the $\pi^+$, is used to identify the $K^{+} \rightarrow
\pi^{+}\pi^{0}$ signal. The following selection criteria are used:

\begin{itemize}
\item[(C-1)] FC events with one or two rings and all rings are
  $e$-like,
\item[(C-2)] one Michel decay electron from the muon produced by
  $\pi^+$ decay,
\item[(C-3)] the reconstructed invariant mass of the $\pi^0$ candidate
  is between 85 and 185\,MeV/$c^{2}$,
\item[(C-4)] the reconstructed momentum of the $\pi^0$ candidate is
  between 175 and 250\,MeV/$c$,
\item[(C-5)] the residual visible energy associated with neither the
  $\pi^0$ nor the $\pi^+$ is low: $E_{res} < $ 12\,MeV for two-ring
  events and $E_{res} < $ 20\,MeV for single-ring events, $E_{res}$ is
  described in latter,
\item[(C-6)] the likelihood for the photon distribution is consistent
  with that expected for signal events: $L_{shape} >$ 2.0 for two-ring
  events and $L_{shape} >$ 3.0 for single-ring events in SK-I/III/IV;
  $L_{shape} >$ 1.0 for SK-II. $L_{shape}$ is explained later.
\item[(C-7)] there is visible energy backwards from the $\pi^0$
  direction consistent with a low momentum $\pi^+$: 10\,MeV $< E_{bk}
  <$ 50\,MeV. A detailed description of $E_{bk}$ is given later.
\end{itemize}

A special $\pi^{0}$ reconstruction algorithm is applied to the
single-ring sample, which was developed for rejecting single-ring
$\pi^{0}$ background from CC $\nu_{e}$ appearance in neutrino
oscillations. The $\pi^{0}$ algorithm forces a fit to the best second
ring by comparing the observed and the expected light patterns under
the assumption of two electromagnetic showers and reconstructs the
invariant mass and momentum of the $\pi^{0}$ candidate. Then (C-3) and
(C-4) can be applied even for the single-ring samples.
 
\begin{figure}[htbp]
\begin{center}
  \includegraphics[width=8cm,clip]{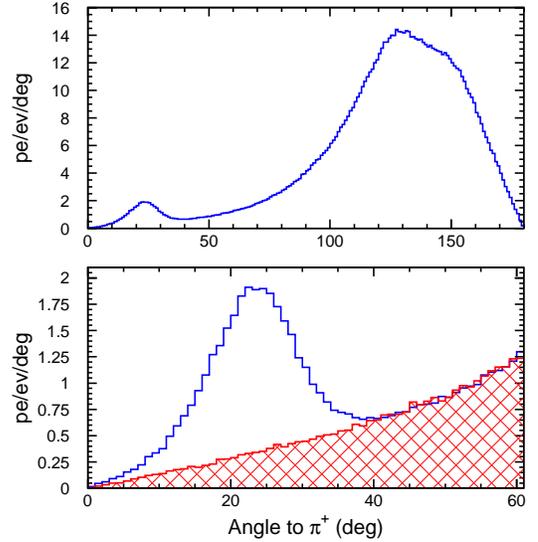}
  \caption
  {\protect \small (color online) Charge distribution as a function of
    angle to the $\pi^{+}$ direction which is defined as opposite the
    reconstructed $\pi^{0}$ direction. The upper figure shows the
    distribution for the signal MC in which $K^{+}$ decays into
    $\pi^{+}$ and $\pi^{0}$. The bump around 23$^\circ$ in the signal
    is made by Cherenkov light of $\pi^{+}$. The lower figure shows
    only the region from 0$^\circ$ to 60$^\circ$ for signal MC (blue)
    and atmospheric $\nu$ MC (hatched red) after the (C-1)-(C-5) criteria are
    required.}
  \label{pi-ang}
\end{center}
\end{figure}

After selecting single $\pi^{0}$ candidates in the signal momentum
region, and requiring a Michel electron, further cuts are applied to
find the tiny Cherenkov light from the $\pi^{+}$. Figure~\ref{pi-ang}
shows the photoelectron distribution versus angle for proton decay
Monte Carlo events. The angle is calculated from the opposite
direction of the reconstructed $\pi^{0}$, which can be assumed as the
$\pi^{+}$ direction. The small bump around 23$^\circ$ comes from
$\pi^{+}$.

\begin{figure}[htbp]
\begin{center}
  \includegraphics[width=4.5cm,clip]{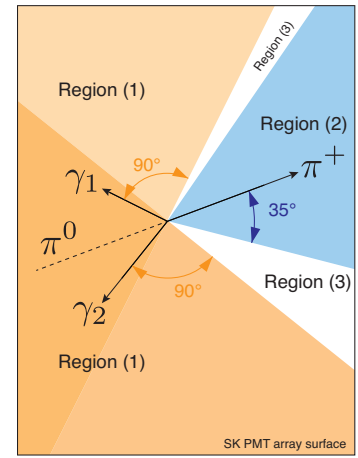}
  \caption
  {\protect \small (color online) An illustration showing each region
    for $K^{+}\rightarrow \pi^{+}\pi^{0}$.  Region (1) is an area
    which is inside of 90$^\circ$ from each $\gamma$ direction, region
    (2) is an area which is inside of 35$^\circ$ from backward of
    reconstructed $\pi^0$ direction, and region (3) is defined as
    residual part.  }
  \label{region}
\end{center}
\end{figure}

The Cherenkov light in an event is then divided into three regions:
(1) a $\pi^{0}$ dominant region, inside of 90$^\circ$ from each
$\gamma$ direction, (2) a $\pi^{+}$ dominant region, inside of
35$^\circ$ from the backwards direction from the reconstructed
$\pi^{0}$ momentum vector, and (3) a residual region. These regions
are illustrated in Fig.~\ref{region}.

\begin{figure}[htbp]
\begin{center}
  \includegraphics[width=8cm,clip]{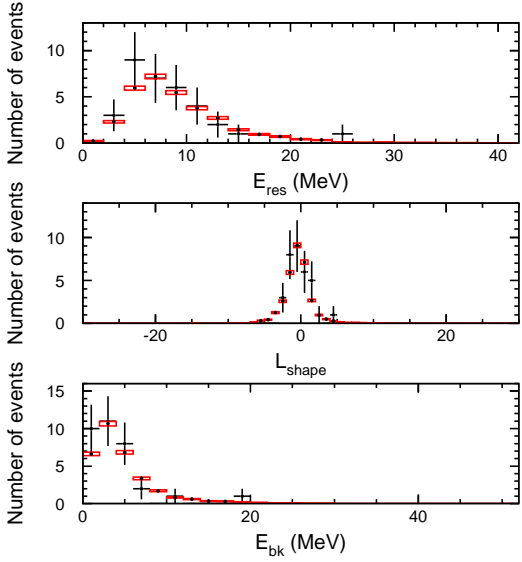}
  \caption
  {\protect \small (color online) $E_{res}$ (upper), $L_{shape}$
    (middle), and $E_{bk}$ (lower) distributions for the two-ring
    sample in SK-I,III, and IV after cuts from (C-1) through
    (C-4). Black crosses correspond to data and red histograms show the
    atmospheric $\nu$ MC normalized by livetime. }
  \label{pipi-data}
\end{center}
\end{figure}

\begin{figure}[htbp]
\begin{center}
  \includegraphics[width=8cm,clip]{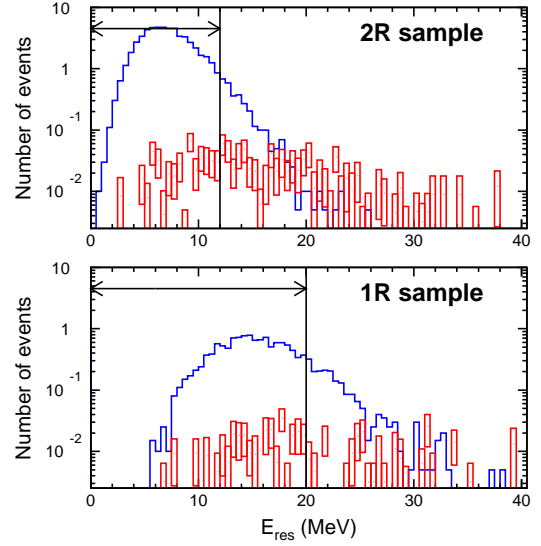}
  \caption
  {\protect \small (color online) $E_{res}$ distributions for the
    two-ring (upper) and single-ring (lower) samples after all
    cuts except (C-5) in all periods. Red histograms are atmospheric
    $\nu$ MC, and blue histograms are proton decay MC,
    respectively. No data remain after cuts.  Arrows in the figures
    show signal regions.}
  \label{eres}
\end{center}
\end{figure}

\begin{figure}[htbp]
\begin{center}
  \includegraphics[width=8cm,clip]{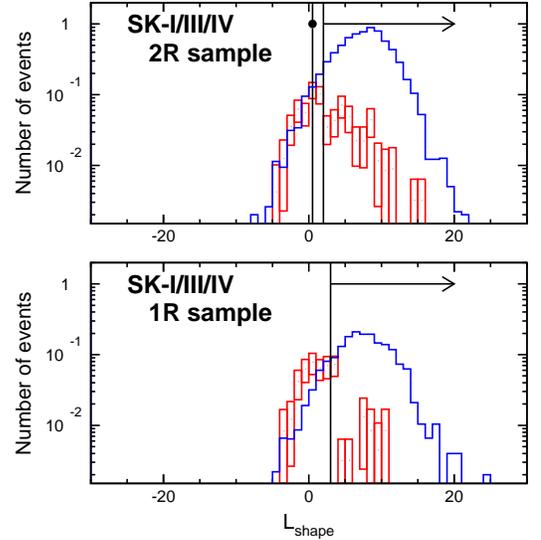}
  \caption
  {\protect \small (color online) $L_{shape}$ distributions for the
    two-ring (upper) and single-ring (lower) samples for
    SK-I,III, and IV, after all cuts except (C-6). Black dots
    correspond to data, red histograms are atmospheric $\nu$ MC
    normalized to the livetime of data, and blue histograms are proton
    decay MC, respectively.  Arrows in the figures show signal
    regions.}
  \label{lshape}
\end{center}
\end{figure}

\begin{figure}[htbp]
\begin{center}
  \includegraphics[width=8cm,clip]{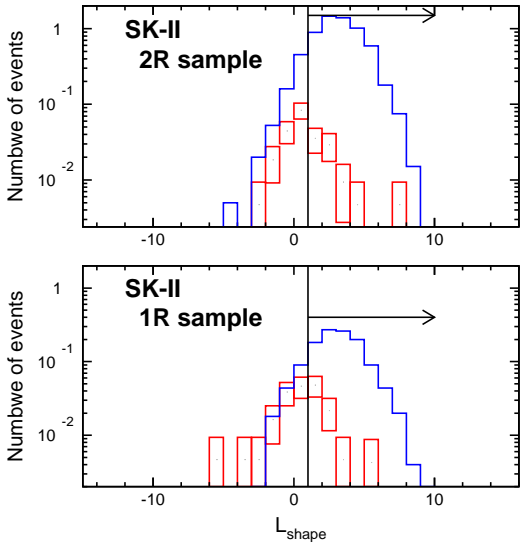}
  \caption
  {\protect \small (color online) $L_{shape}$ distributions for
    the two-ring (upper) and single-ring (lower) samples for SK-II
    after all cuts except (C-6), respectively. Red histograms are
    atmospheric $\nu$ MC, and blue histograms are proton decay MC. No
    data remain after cuts.  Arrows in the figures show signal
    regions.}
  \label{lshape2}
\end{center}
\end{figure}

\begin{figure}[htbp]
\begin{center}
  \includegraphics[width=8cm,clip]{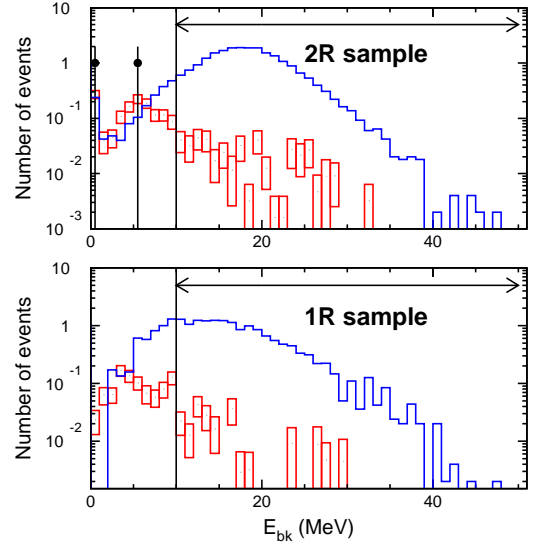}
  \caption
  {\protect \small (color online) $E_{bk}$ distributions for the
    two-ring (uper) and single-ring (lower) samples after all cuts
    except (C-7) in all periods. Black dots correspond to data, red
    histograms are atmospheric $\nu$ MC, and blue histograms are
    proton decay MC, respectively.  Arrows in the figures show signal
    regions.}
  \label{ebk}
\end{center}
\end{figure}

Visible energy sums are calculated for regions (2) and (3) by masking
region (1) to define $E_{bk}$ and $E_{res}$ to be used in criteria
(C-5) and (C-7), respectively. Non-zero $E_{bk}$ is used to identify
the presence of the $\pi^+$.  By requiring low $E_{res}$, we reject
background events with additional final state particles that produce
Cherenkov light. In the case of a single-ring $\pi^0$ candidate, only
the ring direction found by the default reconstruction is masked as
region (1), because the other ring candidate found by the $\pi^{0}$
special algorithm is often found at a large angle from the existing
ring direction due to asymmetric decay, and if the 90$^\circ$ cone of
the additional ring is masked, the $\pi^{+}$ direction may be also
masked. In this case, the missing $\gamma$ may exist in region (2) or
(3), so the cut value for $E_{res}$ is looser than in the two-ring
case. The shape of the angular distribution seen in Fig.~\ref{pi-ang}
is also useful to separate signal and background. Based on
Fig.~\ref{pi-ang}, the expected photoelectrons are generated assuming
signal and background, and a likelihood function ($L_{shape}$) is
calculated for use in (C-6). Then $\pi^{+}$ can be tagged by using
deposited energy ($E_{bk}$) and shape of charge distribution
($L_{shape}$).  The shape of the angular distribution is slightly
different in the single-ring and two-ring cases, and also depends on
photo-coverage, so the cut value is tuned separately for
SK-II. Figure~\ref{pipi-data} shows $E_{res}$, $L_{shape}$, and
$E_{bk}$ distributions of the two-ring sample of SK-I+III+IV data and
the atmospheric $\nu$ MC normalized by livetime, after cuts from (C-1)
through (C-4), with good agreement between data and simulation.
Figures~\ref{eres} through \ref{ebk} show
$E_{res}$, $L_{shape}$, and $E_{bk}$
distributions after all cuts except the cut for itself.

Table~\ref{reduction-pipi} shows background rates per
megaton$\cdot$year exposure, expected events estimated by atmospheric
$\nu$ MC, observed number of events in data, and efficiencies for each
step. Results for SK-II, which has 19\% photo-coverage are separately
shown in the table.  The selection efficiencies, expected numbers of
background, and observed numbers of events for each period are
summarized in Table~\ref{result}. The total expected background for
260~kton$\cdot$year exposure is 0.6 events.  No events are observed in
the SK data. The dominant neutrino interaction modes in the
background are charged current single $\pi$ production (38\%) with low
momentum $\mu$, kaon production (37\%) described in Method 1, and
neutral current multi-$\pi$ production (11\%).

\begin{table*} 
\setlength{\tabcolsep}{7pt} 
\begin{center}
\begin{tabular}{r|rrrr|rrrr}
\hline \hline
           \multicolumn{1}{c}{} &
           \multicolumn{4}{c}{SK-I/III/IV} &
           \multicolumn{4}{c}{SK-II} \\           
Criterion & Bkg. Rate & Exp. Bkg. & Data & Signal Eff. &
             Bkg. Rate & Exp. Bkg & Data & Signal Eff. \\             
\hline
C-1 & 51874.1 & 10940.3 & 10945 & 257
    & 53574.1 & 2635.8  & 2615  & 0.277 \\     
C-2 & 5275.4  & 1112.6  & 1126  & 0.191
    & 5421.1  & 266.7   & 300   & 0.191 \\     
C-3 & 1003.9  & 211.7   & 200   & 0.153 
    & 1266.7  & 62.3    & 56    & 0.137 \\        
C-4 & 225.6   & 47.6    & 42    & 0.131 
    & 244.1   & 12.0    & 10    & 0.114 \\     
C-5 & 171.8   & 36.2    & 37    & 0.120 
    & 178.1   & 8.76    & 7     & 0.101 \\     
C-6 & 10.1    & 2.13    & 2     & 0.098
    & 15.9    & 0.780   & 0     & 0.775 \\ 
C-7 & 2.09    &  0.44   &  0    & 0.087
    & 3.42    &  0.17   &  0    & 0.067 \\          
\hline \hline
\end{tabular}
\end{center}
\caption{\protect \small 
Event rates per Megaton$\cdot$year and expected numbers of event from
atmospheric $\nu$ MC, observed numbers of event in data, and signal
efficiencies estimated from proton decay MC, for method 3. 
SK-I/III/IV with 40\% photo-coverage and SK-II with 19\% are shown separately.  
}
\label{reduction-pipi}
\end{table*}  

The uncertainty for the $\pi^0$ fitter is rather large, 18\%~\cite{t2k},
but is only applied to the fraction of the single-ring $\pi^0$ sample
(19\%) with a final systematic error of 4.5\%. The leading systematic
 uncertainty in the signal efficiency of Method 3 is in the $\pi^+$ 
interaction in water, estimated to be 5\%. Including other
reconstruction errors as listed for Method 1, the total systematic
uncertainty in the signal efficiency of Method 3 is estimated to be
9.5\%. The systematic uncertainty for the background in Method 3 is
estimated to be 29\% based on the uncertainties in atmospheric
neutrino flux and cross section following the same procedure as
described for Method 1.

Compared to our previous publication, Method 3 is improved in
efficiency by the addition of the single-ring $\pi^0$ sample
which occupies 19\% of the selected events in the signal MC.   
The angle cut at 40$^\circ$ applied to calculate the charge sum in region
(2) in the previous paper was rather loose (refer to
Figure~\ref{pi-ang}). The cut value was tuned to 35$^\circ$ by
maximizing $S/\sqrt{N}$ to reduce more background. The new variable
$L_{shape}$ also reduces background. As a result, the efficiency of
Method 3 is increased by 30\% and the background is reduced by 30\%
from the previously published result for SK-I. The efficiencies,
backgrounds, and observed events are summarized in
Table~\ref{result}. The efficiency in SK-IV is larger than the other
periods because of the improvement in efficiency for Michel electron
tagging.

\begin{table}
  \begin{center}
    \begin{tabular}{ll|cccc}
      \hline \hline
                    &                     &  SK-I       &  SK-II      &  SK-III     & SK-IV      \\
      \hline 
Exp.(kton$\cdot$yrs)&                     &  91.7       &  49.2       &  31.9       & 87.3       \\
\hline
Prompt $\gamma$     &  Eff.(\%)           & 7.9$\pm$0.1 &6.3$\pm$0.1  &7.7$\pm$0.1  &9.1$\pm$0.1 \\
                    &  BKG/Mt$\cdot$yr & 0.8$\pm$0.2  & 2.8$\pm$0.5 & 0.8$\pm$0.3 & 1.5$\pm$0.3 \\
                    &  BKG                & 0.08        & 0.14        & 0.03        & 0.13       \\
                    &  OBS     & 0           & 0          & 0           & 0       \\
\hline
$P_{\mu}$ spec.     &  Eff.(\%)          & 33.9$\pm$0.3 &30.6$\pm$0.3 &32.6$\pm$0.3 &37.6$\pm$0.3 \\
                    &  BKG/Mt$\cdot$yr & 2107$\pm$39 & 1916$\pm35$ & 2163$\pm$40    & 2556$\pm$47 \\
                    &  BKG                & 193         & 94.3         & 69.0         & 223.1       \\                    
                    &  OBS                & 177         & 78           & 85           & 226       \\
\hline
$\pi^{+} \pi^{0}$    &  Eff.(\%)          & 7.8$\pm$0.1 &6.7$\pm$0.1 &7.9$\pm$0.1 &10.0$\pm$0.1 \\
                    &  BKG/Mt$\cdot$yr & 2.0$\pm$0.4  & 3.4$\pm$0.6 & 2.3$\pm$0.4 & 2.0$\pm$0.3 \\
                    &  BKG                & 0.18        & 0.17       & 0.09        & 0.18       \\                    
                    &  OBS     & 0           & 0          & 0           & 0       \\
      \hline \hline
    \end{tabular}
  \caption{\protect \small Summary of the proton decay search with
    selection efficiencies and expected backgrounds for each detector
    period.  }
  \label{result}
  \end{center}
\end{table}

\subsection{Lifetime Limit}

In the absence of any excess signal above the background expectation,
we calculate the lower limit on the proton partial lifetime using a
Bayesian method~\cite{bayesian} to incorporate systematic uncertainty.
The calculation method used in our previous publication~\cite{kk} is
applied in this analysis.

For Methods $i=$ 1 and 3, where $n_i$ is the number of candidate
events in the $i$-th proton decay search, the conditional probability
distribution for the decay rate is expressed as:
\begin{eqnarray}
P(\Gamma|n_i) &=& \iiint \frac{e^{-(\Gamma\lambda_i\epsilon_i+b_i)}
(\Gamma\lambda_i\epsilon_i+b_i)^{n_i}}{n_i!} \times \nonumber \\ \nopagebreak
 & & ~~P(\Gamma)P(\lambda_i)P(\epsilon_i)P(b_i)
     \,d\lambda_i\,d\epsilon_i\,db_i
\end{eqnarray}
where $\lambda_i$ is the true detector exposure, $\epsilon_i$ is the
true detection efficiency including the meson branching ratio, and
$b_i$ is the true number of background events. The decay rate prior
probability distribution $P(\Gamma)$ is 1 for $\Gamma \ge 0$ and
otherwise 0.

The prior probability distributions incorporating uncertainties
in detector exposure $P(\lambda_i)$, efficiency $P(\epsilon_i)$, and
background $P(b_i)$, are expressed as:
\begin{eqnarray}
 P(\lambda_i)  &=& \delta(\lambda_i-\lambda_{0,i}) \\
 P(\epsilon_i) &=&  \exp\left[-(\epsilon_i-\epsilon_{0,i})^2/2\sigma_{\epsilon,i}^2\right] \\ 
 & & \mathrm{(0\le\epsilon_i\le1,\; otherwise\; 0)}\nonumber \\
P(b_i) &=& \int^\infty_0\frac{e^{-b'}(b')^{b_{0,i}}}{b_{0,i}!} \exp\left[\frac{-(b_iC_i-b')^2}{2\sigma_{b,i}^2}\right]db' \\
 & & \mathrm{(0\le }b_i\mathrm{, \;otherwise\;0)}\nonumber
\end{eqnarray}

\noindent where $\lambda_{0,i}$ is the estimated exposure,
$\epsilon_{0,i}$ is the estimated efficiency, $b_{0,i}$ is the estimated
number of background events in 500 years MC, $C_i$ is the
MC-to-exposure normalization factor, and $\sigma_{\epsilon,i}$ and
$\sigma_{b,i}$ are the uncertainties in detection efficiency and
background, respectively.

To combine Method 2, the remaining events are divided into three
momentum bins: 200-215\,MeV/$c$, 215-260\,MeV/$c$, and
260-305\,MeV/$c$ and the number of observed events are denoted as
$m_1, m_2, m_3$ instead of $n_i$. Then, the nucleon decay rate
probability, $P(\Gamma|m_1,m_2,m_3)$, is calculated as:
\begin{widetext}
\begin{eqnarray}
P(\Gamma|m_1,m_2,m_3) =  
\iiint\prod^3_{j=1}\frac{e^{-(\Gamma\lambda_j\epsilon_j+b_{shape,j}b)}
(\Gamma\lambda_j\epsilon_j+b_{shape,j}b)^{m_j}}{m_j!}
~P(\Gamma)P(\lambda_j)P(\epsilon_j)P(b)P(b_{shape,j})
\,d\lambda_j\,d\epsilon_j\,db\,db_{shape,j},
\end{eqnarray}
\end{widetext}

\noindent where $j = 1, 2, 3$ corresponds to each momentum bin, $P(b)$
is defined as one for $b > 0$ and otherwise 0, and $\epsilon_{j}$
denotes the efficiency for each bin. The number of background events,
$b_{shape,j}$, is $b_j$ normalized by $b_2$. The uncertainty function
of the background shape $P(b_{shape,j})$ is defined by a Gaussian
function for the 1st and 3rd bin, and a delta function for the 2nd
bin. The uncertainties of the background are estimated to be 7\% and
8\% respectively from the difference of MC models.

We combine all three searches to calculate the lower limit of the
nucleon decay rate, $\Gamma_{\mathrm{limit}}$ as:

\begin{eqnarray}
\mathrm{CL}=\frac{\int^{\Gamma_{\mathrm{limit}}}_{\Gamma=0}\prod^{N=3}_{i=1}P(\Gamma|n_i)\,d\Gamma} {\int^\infty_{\Gamma=0}\prod^{N=3}_{i=1}P(\Gamma|n_i)\,d\Gamma},
\end{eqnarray}
\noindent where $N=3$ is the number of independent search
methods\footnote{for $i=2$, the second search method, $P(\Gamma | n_2)
  \equiv P(\Gamma | m_1,m_2,m_3)$}, and CL is the confidence level,
  taken to be 90\%. The lower lifetime limit is given by:
\begin{eqnarray}
\tau/\mathrm{B}_{p\rightarrow \nu K^+}=\frac{1}{\Gamma_{\mathrm{limit}}}\sum_{i=1}^N[\epsilon_{0,i}\cdot\lambda_{0,i}].
\end{eqnarray}

\noindent The result of the limit calculation combining the three
search methods is:
\begin{eqnarray}
\tau/\mathrm{B}_{p\rightarrow \nu K^+} > 
  5.9 \times 10^{33} \mathrm{years,} \nonumber
\end{eqnarray}
at the 90\% confidence level. If only the results of the low
background searches are used, Methods 1 and 3, the lower limits of
proton lifetime are estimated to be 2.5$\times 10^{33}$ and 2.6$\times
10^{33}$ years, respectively. The lifetime limit from the muon
momentum spectrum fit, Method 2, is $0.8 \times 10^{33}$ years.

\medskip

\section{CONCLUSION}

The proton decay search for $p \rightarrow \nu K^{+}$ was carried out
with 260 kton$\cdot$year exposure, including SK-I, II, III, and IV.
There are several improvements in the analysis and we succeeded to
reduce backgrounds drastically and to increase efficiencies.  However,
we do not find any evidence for proton decay in this exposure,
therefore we have set a limit on the partial lifetime as 5.9$\times
10^{33}$ years, which is more than 2.5 times more stringent than our
previous publication.  The non-observation of proton decay into this
mode constrains, but does not exclude, recent SUSY GUT models.

\section{Acknowledgments}

We gratefully acknowledge the cooperation of the
Kamioka Mining and Smelting Company.
The Super-Kamiokande experiment has been built and operated
from funding by the Japanese Ministry of Education,
Culture, Sports, Science and Technology, the United
States Department of Energy, and the U.S. National Science
Foundation. Some of us have been supported by
funds from the Korean Research Foundation (BK21),
the National Research Foundation of Korea (NRF-20110024009),
the State Committee for Scientific Research
in Poland (grant1757/B/H03/2008/35), the European Union FP7
(DS laguna-lbno PN-284518 and ITN invisibles GA-2011-289442), the Japan
Society for the Promotion of Science, and the National
Natural Science Foundation of China under Grants No.10575056.


\end{document}